\documentstyle[11pt]{article}
\newcommand{\beq}{\begin{equation}}
\newcommand{\eeq}{\end{equation}}
\newcommand{\bea}{\begin{eqnarray}}
\newcommand{\eea}{\end{eqnarray}}
\newcommand{\lpart}{\raise.3ex\hbox{$\stackrel{\leftarrow}{\partial}$}}
\newcommand{\ldr}{\raise.3ex\hbox{$\stackrel{\leftarrow}{\delta^r}$}}
\begin{document}
\thispagestyle{empty}
\vspace{3cm}
\title{\Large{\bf Second Class Constraints\\in a 
Higher-Order Lagrangian Formalism} }
\vspace{2cm}
\author{{\sc I.A. Batalin}\\I.E. Tamm Theory Division\\
P.N. Lebedev Physics Institute\\Russian Academy of Sciences\\
53 Leniniski Prospect\\Moscow 117924\\Russia\\
~\\~{\sc K. Bering}\\Institute of Theoretical Physics\\
Uppsala University\\P.O. Box 803\\S-751 08 Uppsala\\Sweden\\~\\and\\~\\
{\sc P.H. Damgaard}\\The Niels Bohr Institute\\Blegdamsvej 17\\
DK-2100 Copenhagen\\Denmark}
\maketitle
\begin{abstract}
We consider the description of second-class constraints in a 
Lagrangian path integral associated with a higher-order $\Delta$-operator.
Based on two conjugate higher-order $\Delta$-operators, we also propose 
a Lagrangian path integral with $Sp(2)$ symmetry, and describe
the corresponding system in the presence of second-class constraints. 
\end{abstract}
\vspace{4cm}
\begin{flushleft}
NBI-HE-96-60\\UUITP-09/97\\hep-th/9703199
\end{flushleft}

\vfill
\newpage

\noindent
{\em 1.\ Introduction.}~
Recently, we proposed a Lagrangian path integral formalism based
on a $\Delta$-operator that is not restricted to be of second order
\cite{BBD}. The main result was a demonstration that the suggested
gauge-fixed functional integral is independent of the chosen
gauge-fixing function. This generalizes the field--antifield
quantization program \cite{BV}, taken in its more recent
reformulation \cite{BT,BT',BMS}, to the case of gauge symmetries based
on higher-order $\Delta$-operators. It also provides a very concise
way of understanding the mechanism of gauge-independence, even when
restricted to the conventional second-order formalism.
Higher order quantum corrections
to the conventional second-order operator $\Delta^{(2)}$ \cite{BV} are 
expected to arise from the Hamiltonian formalism when operator-ordering
ambuguities are taken into account. A first analysis of this more
general framework was carried out in ref. \cite{BT1}, and it has more
recently been considered from a different point of view \cite{AD,BDA}.
The relevant mathematical structure can be described by a tower
of higher antibrackets \cite{KA}, and their associated algebra \cite{BDA}.

In this paper we shall describe the corresponding construction of
a gauge-fixed Lagrangian path integral in the presence of what 
appropriately may be called second-class constraints \cite{BT2}
within the antibracket formalism.
The possible presence of higher-order terms in the $\Delta$-operator makes
it necessary to reconsider this notion, and we therefore first propose a 
set of conditions the second-class constraints must satisfy. A gauge-fixed
Lagrangian path integral in the presence of these constraints
is then constructed in such a way as to be independent of the chosen
gauge-fixing function. In doing this, we are much helped by a theorem,
proven in ref. \cite{BT1}, concerning the relation between
the second-order operator $\Delta^{(2)}$ and its most general higher-order
local quantum deformations. We finally construct the analogous
Lagrangian path integral (both with and without second-class constraints)
with extended BRST--anti-BRST symmetry.

\vspace{0.7cm}
\noindent
{\em 2.\ Constraints of Second Class.}~ Once the conventional
field-antifield formalism is formulated in terms of the usual antibracket
$(A,B)$, one can consider the problem of reducing the antisymplectic 
$2N$-dimensional manifold down to a physical submanifold of dimension
$2(N - K)$ by means of $2K$ second-class constraints $\Theta^{\alpha}
(\Gamma)$ \cite{BT2}. In the usual case, this proceeds quite analogous
to the Poisson-bracket treatment of second-class constraints in the
Hamiltonian formalism. The crucial ingredients are the following:
For the second-class constraints, the antibracket matrix
\beq
E^{\alpha\beta} ~\equiv~ (\Theta^{\alpha},\Theta^{\beta}) \label{Q}
\eeq
must have an inverse $E_{\alpha\beta}$:
\beq
E_{\alpha\beta}E^{\beta\gamma} ~=~ \delta^{\gamma}_{\alpha} ~.
\eeq
This opens up the possibility of introducing a Dirac antibracket
as the projection \cite{BT2},
\beq
(A,B)_D ~\equiv~ (A,B) - (A,\Theta^{\alpha})E_{\alpha\beta}
(\Theta^\beta,B) ~,\label{Diracbracket}
\eeq
$i.e.$, with the property that the Dirac antibracket vanishes when
taken of any function \mbox{$F=F(\Gamma)$} with any of the constraints:
\beq
(F,\Theta^{\alpha})_D ~=~ 0 ~=~ (\Theta^{\alpha},F)_D ~. \label{thetavanish}
\eeq 
One can also introduce a nilpotent Dirac $\Delta$-operator 
$\Delta_D^{(2)}$, so that the Dirac antibracket (\ref{Diracbracket}) 
equals the failure of $\Delta_D^{(2)}$ to act as a derivation.
This operator commutes\footnote{Here, and in the following, $[\cdot,
\cdot]$ equals the graded supercommutator: $[A,B] = AB - (-1)^{
\epsilon_A\epsilon_B}BA$.} with the second-class constraints:
\beq
[\Delta_D^{(2)},\Theta^{\alpha}] ~=~ 0 ~.\label{delta2com}
\eeq
With these ingredients, the corresponding Lagrangian path integral
formulation can be carried out very analogous to the case
without second-class constraints \cite{BT2}.

Our aim is to generalize this construction to the case of a higher-order
$\Delta$-operator. We do this in the framework of
local quantum deformations \cite{BT1}. This means that we are going to infer
the proper meaning of second-class constraints in the higher-order
formalism from the corresponding second-order formulation. We thus start
with a second-order operator $\Delta^{(2)}$. 
Second-class constraints can now be treated
\cite{BT2} by means of functions $\Theta^{(2)\alpha}(\Gamma)$
and a Dirac $\Delta$-operator $\Delta^{(2)}_D$ satisfying the above
conditions. We next apply the deformation theorem to
construct, with the help of a transformation 
\mbox{$U=U(\Gamma,\partial)$}, 
a general higher-order nilpotent Dirac 
$\Delta$-operator and the associated second-class constraints:
\begin{eqnarray}
\Delta_D &~=~& U^{-1} \Delta^{(2)}_D U ~=~ \Delta^{(2)}_D + {\cal O}(
\hbar) \cr
\Theta^{\alpha} &~=~& U^{-1} \Theta^{(2)\alpha} U ~=~ \Theta^{(2)\alpha}
+ {\cal O}(\hbar) ~.
\end{eqnarray}
One unusual consequence of this construction is that the second-class
constraints 
\mbox{$\Theta^{\alpha} = \Theta^{\alpha}(\Gamma,\partial/
\partial\Gamma)$} 
in general are {\em operators}. Nevertheless, by construction,
\beq
[\Delta_D,\Theta^{\alpha}] ~=~ 0 ~~,~~~~~
[\Theta^{\alpha},\Theta^{\beta}] ~=~ 0 ~,\label{thetaprop}
\eeq
the latter of these relations of course trivially being satisfied for
the $\Theta^{(2)\alpha}$'s. Because the constraints are now operators,
we are led to require that physical functions and
operators \mbox{$F=F(\Gamma,\partial/\partial\Gamma)$} commute
with the second-class constraints: $[F,\Theta^{\alpha}] ~=~0$.

A particularly interesting subset of deformed theories will consist of
those for which the deformation is consistent with the original
second-class constraints. In this case $[U,\Theta^{(2)\alpha}] = 0$. The
second-class constraints of the higher-order formalism then remain 
unchanged, and they are in particular still functions. 
Deformations based on transformations $U$ which
depend only on $\Gamma^A$ and projected differentiations
${\lpart}_B{\cal P}^B_{~~A}$ with 
\beq
{\cal P}^A_{~~B} ~\equiv~  \delta^A_{~~B} - (\Gamma^A,\Theta^{\alpha})
E_{\alpha\beta}(\Theta^{\beta}{\lpart}_B) ~\label{proj}
\eeq
are guaranteed to fall into this class. One could restrict oneself to
the set of higher-order $\Delta$-operators that arise from such deformations, 
in which case the following analysis would be much simplified.
But we shall here consider the more general framework in which the
second-class constraints are allowed to be operators.

Given solutions $W^{(2)}$ and $X^{(2)}$ of the Master Equations
\beq
\Delta^{(2)}_D e^{\frac{i}{\hbar}W^{(2)}} ~=~ 0 ~~~,~~~~~
\Delta^{(2)}_D e^{\frac{i}{\hbar}X^{(2)}} ~=~ 0 ~,\label{2ndme}
\eeq
we immediately have the solutions
\beq
e^{\frac{i}{\hbar}W} ~=~ U^{-1} e^{\frac{i}{\hbar}W^{(2)}} ~~~,~~~~~
e^{\frac{i}{\hbar}X} ~=~ U^{-1} e^{\frac{i}{\hbar}X^{(2)}} 
\eeq
to the higher-order Master Equations
\beq
\Delta_D e^{\frac{i}{\hbar}W} ~=~ 0 ~~~,~~~~~
\Delta_D e^{\frac{i}{\hbar}X} ~=~ 0 ~.\label{higherme}
\eeq
These, in turn, can be expanded \cite{AD,BDA} in terms
of the higher antibrackets $\Phi^{k}_{\Delta_{D}}$, {\em viz.,}
\beq
e^{-\frac{i}{\hbar}W}\Delta_D e^{\frac{i}{\hbar}W} ~=~
\sum_{k=0}^{\infty}\left(\frac{i}{\hbar}\right)^k 
\Phi^k_{\Delta_{D}}(W,\ldots, W) ~.
\eeq
We can define the $k$th Dirac antibracket
in term of $k$ nested commutators acting on unity \cite{KA,BDA}:
\beq
\Phi^k_{\Delta_{D}}(A_1,\ldots, A_k) ~\equiv~ [[\cdots[\Delta_D,A_1],
\cdots,A_k]~1 ~.
\eeq
With this definition we can even consider antibrackets of operators.
If these operators (including functions) all commute, then all
standard properties of these higher antibrackets (such as their
strongly homotopy Lie algebra) are preserved \cite{BDA}. In particular, 
we can consider the Dirac antibrackets with one (or more) of the 
second-class constraints
inserted, all other slots being occupied by physical (see above)
and mutually commuting operators. By virtue of eq. (\ref{thetaprop}), all 
these higher Dirac antibrackets vanish identically:
\beq
\Phi^k_{\Delta_{D}}(A_1,\ldots,A_{i-1},\Theta^{\alpha},A_{i+1},\ldots,
A_k) ~=~ 0 ~.
\eeq
This is the obvious generalization to higher order of the Dirac
antibracket property (\ref{thetavanish}). It means that in the algebra of
higher antibrackets based on $\Delta_D$, the $\Theta^{\alpha}$'s are
zero operators in the strong sense.

Consider now the appropriate prescription for the Lagrangian path integral
in the presence of these second-class constraints. In the second-order
formalism, a solution to this problem was given in ref. \cite{BT2}.
In the general higher-order formalism, the functional measure is such that 
the Dirac operator $\Delta_D$ is symmetric, 
\beq
\Delta_D ~=~ \Delta_D^T ~. 
\eeq
Here transposition is defined, for an arbitrary
operator ${\cal A}$, by
\beq
\int d\mu~ F \left(\rho_{\Theta} {\cal A} G\right) ~=~ 
(-1)^{\epsilon_F\epsilon_{\cal A}}\int d\mu~ (
{\cal A}^T F) \left( \rho_{\Theta} G \right)~,
\eeq
where, because $\rho_{\Theta}$ 
in general will be operator valued, 
its position in the integrand is fixed. Here 
\mbox{$d\mu = d\Gamma d\lambda\rho(\Gamma)$} 
is an overall measure, without any reference to
the second-class constraints. In the case of a second-order $\Delta$-operator,
we combine $\rho(\Gamma)$ and $\rho^{(2)}_{\Theta}(\Gamma)$ into the the
antisymplectic Dirac measure 
\mbox{$\rho_D(\Gamma) = \rho(\Gamma)\rho^{(2)}_{\Theta}(\Gamma)$}. 
Note that in the present context
\beq
\rho_{\Theta} ~=~ U^{-1}\rho_{\Theta}^{(2)} U ~=~ 
\rho_{\Theta}^{(2)} + {\cal O}(\hbar) ~,
\eeq
where $\rho_{\Theta}^{(2)}$ corresponds to the  
second-order formalism with 2nd class constraints $\Theta^{(2)\alpha}$.
Terms proportional to $\hbar$ and higher will in general be operator valued.

In the second-order case \cite{BT2}, where indeed the $\Delta_D$-operator is
symmetric according to the above definition, the gauge-fixed path integral 
can be represented in the form
\beq
Z^D_X ~=~ \int d\Gamma d\lambda~\rho_D(\Gamma)\prod_{\alpha} 
\delta(\Theta^{(2)\alpha})e^{\frac{i}{\hbar}(W^{(2)} + X^{(2)})} ~.
\label{2nd2nd}
\eeq
Remarkably, a closed-form expression for the gauge
fixing $X$ can be found in this second-order case\footnote{Up to terms 
proportional to $\lambda^*$, which are not of our concern here. See,
$e.g.$, ref. \cite{BMS}.} This is the generalization 
of the expression given in ref. \cite{BT'} (see also the discussion
in ref. \cite{K}). To find it, we first introduce 
gauge-fixing 
functions $G_{a}$ in involution with respect to the Dirac antibracket,
\beq
(G_a,G_b)_D ~=~ G_c U^{c}_{ab} ~.
\eeq
Now consider a solution to the Master Equation for $X^{(2)}$ (\ref{2ndme}) of
the form \mbox{$X^{(2)} = G_a\lambda^a + i\hbar H + \ldots$}, 
where the omitted terms are at least proportional to $\lambda^*$. 
Requiring the Master Equation (\ref{2ndme}) for $X$ to be satisfied, 
one finds the explicit solution
\beq
\exp[-H] ~=~ \sqrt{{\mbox{\rm sdet}}(F^a,G_b)_D \ J_D\  \rho_D^{-1}} ~,
\label{H}
\eeq
where 
\mbox{$J_D ~=~ {\mbox{\rm sdet}} (\bar{\Gamma}^A {\lpart}_B )$}
is the Jacobian of the change of coordinates
\beq
\Gamma^A ~\longrightarrow~
\bar{\Gamma}^A \equiv \{F^a;G_a; \Theta^{(2)\alpha} \}~.
\eeq
The solution $H$ as defined in 
eq.\ (\ref{H}) is independent of 
the choice of $F^a$ on the surface $G_a=0$.

Comparing with the second-order case \cite{BT2}, we now propose
the following gauge-fixed path integral in the general case\footnote{
The $\delta$-function of an operator is given meaning in terms of a
suitable representation and its formal Taylor expansion.}:
\beq
Z^D_X ~=~ \int d\mu~ e^{\frac{i}{\hbar}W}\prod_{\alpha}
\delta(\Theta^{\alpha}) \rho_{\Theta} e^{\frac{i}{\hbar}X} ~.\label{2ndhigher}
\eeq
Here both $W$ and $X$ satisfy the appropriate higher-order Master Equations
(\ref{higherme}). The transformation $U$ should be orthogonal.
Using the technique of ref. \cite{BBD}, and the condition (\ref{delta2com}),
one verifies the crucial property that this path integral is independent 
of the gauge fixing $X$. 

The integral (\ref{2ndhigher}) is invariant under changes of bases of
the second-class constraints $\Theta^{\alpha}$. All our defining
equations, including that of the Dirac antibracket itself, have been formulated
in the strong sense. Although these equations are unstable under
reparametrizations of the constraints, there is no need to consider
the corresponding set of weak equations, due to the presence of the
$\delta(\Theta^{\alpha})$-factor in the integrand of (\ref{2ndhigher}).
This aspect is completely analogous to the situation in the Hamiltonian path
integral with second-class constraints.

Despite its apparently asymmetric formulation in eq.\ (\ref{2ndhigher})
we also observe that
\beq 
Z^D_X ~=~ \int d\mu~ e^{\frac{i}{\hbar}W}\prod_{\alpha}
\delta(\Theta^{\alpha}) \rho_{\Theta} e^{\frac{i}{\hbar}X} 
~=~\int d\mu~ e^{\frac{i}{\hbar}X}\prod_{\alpha}
\delta(\Theta^{\alpha}) \rho_{\Theta} e^{\frac{i}{\hbar}W} ~,
\eeq
as a result of 
\mbox{$(\Theta^{\alpha})^T = \Theta^{\alpha}$} and 
\mbox{$\rho_{\Theta}^T =\rho_{\Theta}$}. 
Thus, as in the formulation without second-class constraints, only
boundary conditions distinguish the ``action'' $W$ from its ``gauge
fixing'' $X$. 

\vspace{0.7cm}
\noindent
{\em 3.\ The Sp(2) Construction.}~ All of the above considerations can
be generalized to the case of triplectic quantization, which is based
on an $Sp(2)$-symmetric formulation that includes both BRST and
anti-BRST symmetries \cite{BM,BMS,ND,BM1}. Consider
first the case where there are no second-class constraints. We need
to provide the appropriate generalization of triplectic quantization
to the case of higher-order $\Delta$-operators. Fortunately, this
generalization can rather easily be inferred from our earlier paper
\cite{BBD}. Introduce a pair of $\Delta$-operators
\beq
\Delta^a_{\pm} ~\equiv~ \Delta^a \pm \frac{i}{\hbar}V^a ~,\label{deltapm}
\eeq
with the following algebra (curly brackets indicate symmetrization in
the indices):
\beq
\Delta^{\{a}_{\pm}\Delta^{b\}}_{\pm} ~=~ 0 ~.
\eeq
The operators $\Delta^a$ are of second order or higher, while the operator
$V^a$ is of first order only. 
Note also the split in terms of powers of $\hbar$. From
the point of view of quantum deformations of the usual second-order
triplectic formalism, the $\Delta^a$-operators of eq. (\ref{deltapm}) contain
infinite expansions in $\hbar$, starting with the classical terms. 

Let there now be given a functional measure $d\mu \equiv d\Gamma d\lambda
\rho(\Gamma)$ on a $6N$-dimensional triplectic space of fields.
The operators $\Delta^a_+$ and $\Delta^a_-$ are required to be conjugate
to each other in the sense that
\beq
\int d\mu~ F\Delta^a_{\pm} G ~=~ (-1)^{\epsilon_F} \int d\mu~
\left(\Delta^a_{\mp} F\right) G ~. \label{sp2conj}
\eeq
We define the Master Equations for $W$ and $X$ to be
\beq
\Delta^a_+ e^{\frac{i}{\hbar}W} = 0 ~,~~~~~ \Delta^a_-
e^{\frac{i}{\hbar}X} = 0 ~,\label{sp2mastereq}
\eeq
and then construct on the basis of their solutions the path integral
\beq
Z_X ~=~ \int d\mu~ e^{\frac{i}{\hbar}[W + X]} ~.\label{sp2part}
\eeq
The description of the Master Equations (\ref{sp2mastereq}) in terms of
higher antibrackets, and the $Sp(2)$ generalization of the algebra these
higher antibrackets satisfy are decribed in ref. \cite{BDA}.

Consider now operators $\Delta^a_{\pm}$ and solutions $X$ to the Master
Equation (\ref{sp2mastereq}) belonging to the class for which
\beq
e^{\frac{i}{\hbar}X'} = e^{\varepsilon_{ab}\frac{\hbar}{i}[\Delta^a_-,
[\Delta^b_-,\Phi]]}e^{\frac{i}{\hbar}X}
\eeq
is a {\em maximal} deformation of $X$ which preserves (\ref{sp2mastereq})
\cite{BLT,BMS}. The partition function (\ref{sp2part}) is unchanged under
this maximal deformation:
\begin{eqnarray}
\delta Z_X &~=~& \left(\frac{\hbar}{i}\right)
\int d\mu~ e^{\frac{i}{\hbar}W}\varepsilon_{ab}
[\Delta^a_-,[\Delta^b_-,\Phi]]e^{\frac{i}{\hbar}X} \cr
&~=~& \left(\frac{\hbar}{i}\right)\varepsilon_{ab}\int d\mu~  
\left[\left(\Delta^a_+ e^{\frac{i}{\hbar}W}\right)\left([\Delta^b_-,\Phi]
e^{\frac{i}{\hbar}X}\right) + [\Delta^b_-,\Phi]\Delta^a_- e^{\frac{i}{\hbar}
X}\right] \cr
&~=~& 0 ~, \label{sp2indep}
\end{eqnarray}
as follows by using the Master Equations (\ref{sp2mastereq}). The gauge
freedom is parametrized in terms of one bosonic function $\Phi$.

As in the case without $Sp(2)$ symmetry \cite{BBD}, we can also here
understand the independence of the gauge-fixing function $X$ from the
existence of a BRST symmetry. Introduce BRST operators
$\sigma^a_W$ and $\sigma^a_X$ by \cite{BDA}
\begin{eqnarray}
\sigma^a_W F &~\equiv~& \frac{\hbar}{i}e^{-\frac{i}{\hbar}W}[\Delta^a_+,
F] e^{\frac{i}{\hbar}W} ~,\cr
\sigma^a_X F &~\equiv~& \frac{\hbar}{i}e^{-\frac{i}{\hbar}X}[\Delta^a_-,F]
e^{\frac{i}{\hbar}X} ~.
\end{eqnarray}
These generalize in an obvious way the $Sp(2)$-covariant ``quantum BRST
operator''\footnote{For the case without $Sp(2)$ symmetry, see ref. \cite{H}.}
derived in ref. \cite{DDB} to the case of a higher-order formalism.
The Master Equations for $W$ and $X$ are preserved under transformations
$W \to W + \varepsilon_{ab}\sigma^a_W\sigma^b_W F$ and
$X \to X + \varepsilon_{ab}\sigma^a_X\sigma^b_X F$, respectively. Thus,
the statement of gauge independence of the path integral (\ref{sp2part})
can, equivalently, be rephrased as $Z_X = Z_{X+\delta X}$ with
$\delta X = \varepsilon_{ab}\sigma^a_X\sigma^b_X F$. The formalism is
symmetric under exchanges of $X$ with $W$; only boundary conditions
stipulate which part will play the r\^{o}le of gauge fixing (here chosen
to be $X$).

It is straightforward to derive, using eq. (\ref{sp2conj}), the expected
properties of the above BRST operators. For example, if we define BRST
invariant operators $G$ by $\sigma^a_W G = 0$, then their expectation
values will not depend on X:
\beq
\langle G\rangle_{X+\delta X} - \langle G\rangle_X ~=~ 0 ~,
\eeq
and we similarly find $\langle \sigma_W^a F \rangle = 0$ for any $F$. Note
that, as expected, the $\sigma^a$'s satisfy the $Sp(2)$-algebra
$\sigma^{\{a}\sigma^{b\}} = 0$ \cite{BDA}.

Let us now finally consider the analogue of second-class constraints
in this higher-order triplectic formalism. We can proceed essentially
as in the previous case, and shall therefore be brief. 
In the second-order formalism, one can define the matrix of 
second-class constraints $\Theta^{\alpha}$ by \cite{BM1}
\beq
E^{\alpha\beta a} ~\equiv~ (\Theta^{\alpha},\Theta^{\beta})^a ~,
\eeq  
where an inverse matrix, in the following sense, is assumed to exist:
\beq
E^{\alpha\beta a}Y^c_{ab\beta\gamma} ~=~ \delta^c_b\delta^{\alpha}_{\gamma} ~.
\eeq
A direct triplectic counterpart of the Dirac antibracket is then \cite{BM1}
\beq
(A,B)^a_D ~\equiv~ (A,B)^a - (A,\Theta^{\beta})^bY^a_{bc\beta\gamma}
(\Theta^{\gamma},B)^c ~,\label{sp2dirac}
\eeq 
and similarly one can introduce a nilpotent Dirac $\Delta^{(2)a}_D$-operator
\cite{BM1}. In the higher order formalism, apply the deformation theorem
\cite{BT2} as before, and introduce, by means of a
transformation $U(\Gamma,\partial)$,
\begin{eqnarray}
\Delta^a_{D\pm} &~=~& U^{-1} \Delta^{(2)a}_{D\pm} U ~=~ 
\Delta^{(2)a}_{D\pm} + {\cal O}(\hbar) \cr
\Theta^{\alpha} &~=~& U^{-1} \Theta^{(2)\alpha} U ~=~ \Theta^{(2)\alpha}
+ {\cal O}(\hbar) ~.
\end{eqnarray}
Again, by construction,
\beq
[\Delta_{D\pm},\Theta^{\alpha}] ~=~ 0 ~~,~~~~~
[\Theta^{\alpha},\Theta^{\beta}] ~=~ 0 ~.\label{sp2thetaprop}
\eeq
Solutions of the Master Equations (\ref{sp2mastereq}), with $\Delta^a_{\pm}$
replaced by $\Delta^a_{D\pm}$,
can then again be expressed in terms of the solutions to the second-order
equations:
\beq
e^{\frac{i}{\hbar}W} ~=~ U^{-1} e^{\frac{i}{\hbar}W^{(2)}} ~~~,~~~~~
e^{\frac{i}{\hbar}X} ~=~ U^{-1} e^{\frac{i}{\hbar}X^{(2)}} ~. 
\eeq
They can also be expanded in terms of $Sp(2)$-covariant higher antibrackets
(see ref. \cite{BDA}) $\Phi^{(k)a}_{\Delta_{\pm}^a}$ which satisfy
an $Sp(2)$ generalization of the usual strongly homotopy Lie algebra.
Using the definition in terms of nested commutators acting on unity
\cite{BDA}, also these higher antibrackets can be taken with operators 
as entries. For physical (in the sense given above) and mutually 
commuting operators, we find the expected generalization:
\beq
\Phi^{(k)a}_{\Delta_{D\pm}}(A_1,\ldots,A_{i-1},\Theta^{\alpha},A_{i+1},\ldots,
A_k) ~=~ 0 ~.
\eeq
The triplectic second-class constraints reduce the dimension of the
``triplectic manifold'' to $6(N - K)$. The obvious $Sp(2)$-generalization 
of the Lagrangian path integral in the presence of these constraints
is precisely as in (\ref{2ndhigher}). The only difference is that $W$
and $X$ satisfy the appropriate $Sp(2)$ Master Equations (\ref{sp2mastereq}).
It is straightforward to verify the gauge independence of this path integral,
exactly as in eq. (\ref{sp2indep}).

\vspace{0.8cm}
\noindent
{\sc Acknowledgement:}~I.A.B. would like to thank the Niels Bohr Institute 
and Uppsala University for the warm hospitality
extended to him there.
The work of I.A.B. and P.H.D. is partially supported by grant INTAS-RFBR
95-0829.
The work of I.A.B. is also supported by grants INTAS 93-2058, INTAS 93-0633,
RFBR 96-01-00482, RFBR 96-02-17314, and NorFA 97.40.002-O.


\vspace{0.3cm}




\begin{thebibliography}{999}
\bibitem{BBD}I.A. Batalin, K. Bering and P.H. Damgaard, Phys. Lett.
{\bf B389} (1996) 673.
\bibitem{BV}I.A. Batalin and G.A. Vilkovisky, Phys. Lett. {\bf 102B}
(1981) 27; Phys. Rev. {\bf D28} (1983) 2567 [E: {\bf D30} (1984) 508];
Nucl. Phys. {\bf B234} (1984) 106.
\bibitem{BT}I.A. Batalin and I.V. Tyutin, Mod. Phys. Lett. {\bf A8}
(1993) 3673; 
\bibitem{BT'}I.A. Batalin and I.V. Tyutin, Mod. Phys. Lett. {\bf A9} 
(1994) 1707. 
\bibitem{BMS}I.A. Batalin, R. Marnelius and A.M. Semikhatov, Nucl. Phys.
{\bf B446} (1995) 249.
\bibitem{BT1}I.A. Batalin and I.V. Tyutin, Int. J. Mod. Phys. {\bf A9}
(1994) 517.
\bibitem{AD} J. Alfaro and P.H. Damgaard, Phys. Lett. {\bf B369} (1996) 289.
\bibitem{BDA} K. Bering, P.H. Damgaard and J. Alfaro, Nucl. Phys. {\bf 
B478} (1996) 459.
\bibitem{KA}J.-L. Koszul, Ast\'{e}risque, hors serie (1985) 257.\newline
F. Akman, q-alg/9506027.
\bibitem{BT2}I.A. Batalin and I.V. Tyutin, Int. J. Mod. Phys. {\bf A8}
(1993) 2333.
\bibitem{K}O.M. Khoudaverdian, hep-th/9508174.
\bibitem{BM}I.A. Batalin and R. Marnelius, Phys. Lett. {\bf B350} (1995) 44.
\bibitem{ND}A. Nersessian and P.H. Damgaard, Phys. Lett. {\bf B355}
(1995) 150.
\bibitem{BM1}I.A. Batalin and R. Marnelius, Nucl. Phys. {\bf  B465}
(1996) 521.
\bibitem{BLT}I.A. Batalin, P.M. Lavrov and I.V. Tyutin,
J. Math. Phys. {\bf 32} (1990) 2513.
\bibitem{H}M. Henneaux, Nucl. Phys. B (Proc. Suppl.) 18A (1990) 47.
\bibitem{DDB}P.H. Damgaard, F. De Jonghe and K. Bering, Nucl. Phys.
{\bf B455} (1995) 440.

\end{thebibliography}
 \end{document}